\documentclass{jfm}

\usepackage[english]{babel}
\usepackage[utf8x]{inputenc}
\usepackage[T1]{fontenc}

\usepackage{amsmath}
\usepackage{graphicx}
\usepackage{subcaption}
\usepackage{booktabs}
\usepackage{float}
\usepackage{graphicx}
\usepackage{epstopdf, epsfig}
\usepackage{flushend}
\usepackage[dvipsnames]{xcolor}
\usepackage{latexsym}
\usepackage{caption}
\usepackage{color,soul}
\usepackage{physics}
\usepackage[normalem]{ulem}

 \usepackage{physics}
 \usepackage{comment}

\graphicspath{{figures/}}
\usepackage[colorlinks=true, allcolors=blue]{hyperref}
\usepackage{xcolor}
\usepackage{pgfplots}
\usepackage{bm}
\usepackage{todonotes}

\newcommand{\St}{\mbox{\it St}}

\newcommand*{\starnr}{\stepcounter{equation}\tag{\theequation}}


\definecolor{dg}{rgb}{0, 0.125, 0.376}

\title{Bispectral decomposition and energy transfer in a turbulent jet}

\author{Akhil Nekkanti\aff{1}, Ethan Pickering\aff{2},  Oliver T. Schmidt\aff{3}, Tim Colonius\aff{1}}

\affiliation{
\aff{1} Division of Engineering and Applied Science, California Institute of Technology, Pasadena, CA, USA
\aff{2} Bayer Crop Science, Boston, MA, USA
\aff{3} Department of Mechanical and Aerospace Engineering, University of California San Diego, La Jolla, CA, USA
}

\begin{document}
    \maketitle
    \begin{abstract}

We employ bispectral mode decomposition (BMD) to investigate coherent triadic interactions and nonlinear energy transfer in a subsonic turbulent jet. BMD extracts the flow structures corresponding to the dominant triadic interactions. We find a strong triadic correlation among the Kelvin-Helmholtz wavepacket, its conjugate, and the streaks. The most energetic streaks occur at the azimuthal wavenumber $m=2$, with the dominant contributing azimuthal wavenumber triad being $[m_1,m_2,m_3]=[1,1,2]$. The spectral energy budget reveals that nonlinear triadic interactions represent an energy loss to the streaks. Analysis across a wide range of frequencies and azimuthal wavenumbers identifies the direction of nonlinear energy transfer and the spatial regions where these transfers are most active. The turbulent jet exhibits a forward energy cascade in a global sense, though the direction of energy transfer varies locally. In the shear layer near the nozzle exit, triadic interactions between relatively smaller scales are dominant, leading to an inverse energy cascade. Farther downstream, beyond the end of the potential core, triadic interactions between larger scales dominate, resulting in a forward energy cascade.
    \end{abstract}

\section{Introduction}

Triadic interactions are fundamental to the transfer of energy in turbulent flow. These interactions arise from the quadratic nonlinearity of the convective term in the Navier-Stokes equation. A triad is a combination of three wave vectors that form a triangle, represented as 
\begin{align}
    \mathbf{k}_j \pm \mathbf{k}_k \pm \mathbf{k}_l = 0 \label{eqn:zerosum1}\\
    f_1 \pm f_2 \pm f_3 = 0,
\end{align}
where $\mathbf{k}_j,\mathbf{k}_k,\mathbf{k}_l$,  denote wavenumber vectors, and ${f_j, f_k, f_l}$ denote frequencies. Triadic interactions and their role in energy transfer has been investigated in homogeneous turbulence \citep{domaradzki1990local,waleffe1992nature,moffatt2014note}. 
These studies categorize triadic interactions as either \emph{local} or \emph{nonlocal}. Local interactions refer to energy transfer between wavelengths of similar scales, while nonlocal interactions involve energy transfer between length scales that differ. For homogeneous turbulence, \citet{domaradzki1990local} demonstrated that nonlocal triadic interactions are primarily responsible for energy transfer. Triadic interactions have also been examined in other types of flows, such as oceanographic flows \citep{phillips1960dynamics, mccomas1977resonant, garrett1975space} and transitional flows \citep{craik1971non, smith1987resonant, mankbadi1993critical}. However, triadic interactions and their associated energy transfer in turbulent jets have yet to be explored in detail. 


Coherent structures in turbulent jets are the primary sources of aft-angle noise \citep{jordan2013wave} and have been investigated for many years. Pioneering work by \citet{crow1971orderly} identified the presence of large-scale coherent structures in turbulent jets, and subsequent researchers \citep{michalke1977instability, crighton1976stability, tam1980radiation, tam1984sound, crighton1990shear} modeled these coherent structures, with varying degrees of success, as linear stability solutions of turbulent mean flow.  In more recent work, the resolvent approach, pioneered by \citet{McKeonJFM2010} for boundary layers, has clarified the fundamental linear mechanisms leading to amplification of disturbances in turbulent jets \citep{SchmidtJFM2018,pickering2020lift}.  These mechanisms include Kelvin-Helmholtz (KH) wavepackets \citep{crighton1976stability}, Orr wavepackets \citep{tissot2017wave,SchmidtJFM2018}, and streaks \citep{Nogueira2019Streaks,pickering2020lift}. The resolvent framework also shows that the optimal nonlinear forcing for the Orr wavepackets and streaks occurs extensively throughout the jet shear layers and fully developed region, whereas the KH wavepackets are convectively amplified from spatially compact regions much closer to the nozzle exit.  These interpretations are also consistent with local (spatial) stability analysis, where KH instability is modal, whereas streaks and Orr wavepackets are nonmodal.  

In jets, especially those with initially turbulent boundary layers, much less is understood about nonlinear mechanisms and the presence of self-sustaining processes involving the dominant amplification mechanisms.  In boundary layers, nonlinear interactions are crucial to the self-sustaining near-wall cycle \citep{waleffe1997self, jimenez2013linear, bae2021nonlinear}, supplying energy to rolls and extracting energy from streaks \citep{waleffe1998streamwise, dessup2018self}.  A few studies have explored how nonlinearities specifically contribute to coherent structures in jets \citep{wu2009low, sandham2008nonlinear, suponitsky2010linear, wu2019nonlinear, zhang2023nonlinear}. For instance, \citet{zhang2023nonlinear} demonstrated that streaks are tied to helical KH modes in a weakly nonlinear expansion of a weakly nonparallel jet flow.  Nonlinear dynamics of coherent structures have also been explored by acoustic (and other) forcing near the nozzle. These flows exhibit tonal responses indicative of strong nonlinear resonant interactions. Forcing the jet at the azimuthal wavenumbers $\pm m$ triggers nonlinear interactions that result in mean flow distortion \citep{cohen1987evolution,long1992controlled,corke1993resonance}. \citet{samimy2007active} found that forcing helical modes significantly shortens the jet potential core and substantially enhances entrainment. Other studies \citep{husain1995experiments,paschereit1995experimental,broze1996transitions,shaabani2019vortex,nekkanti2024nonlinear} have examined the response of initially laminar jets. \citep{broze1994nonlinear,broze1996transitions} studied how forcing an initially laminar jet at different frequencies and amplitudes leads to periodic and chaotic patterns as the flow transitions to turbulence. 

An outstanding question is whether, in a natural (unforced) initially turbulent jet, are there important nonlinear interactions amongst the dominant coherent structures, and, if so, which structures, and at which frequencies and azimuthal mode numbers?  Do interactions differ in the shear layer and fully developed regions? To answer these questions, we use bispectral mode decomposition (BMD), a technique recently proposed by \cite{schmidt2020bispectral}. BMD extends classical bispectral analysis to multi-dimensional and multivariate data. The classical bispectrum, or its normalized form, the bicoherence, measures quadratic phase coupling at a single spatial point, as detailed by \citet{hasselmann1963bispectra, brillinger1965introduction, jenkins1965survey}. The bispectrum has been used to analyze homogeneous turbulence flows \citep{yeh1973spectral, lii1976bispectral, herring1980theoretical} and laminar-turbulent transitions \citep{corke1989resonant, hajj1992subharmonic, hajj1993fundamental}. BMD facilitates the extraction of spatial flow structures associated with triadic interactions by maximizing the spatially integrated bispectrum. BMD has previously been applied to various flow configurations, including laminar-turbulent transition on a flat plate \citep{goparaju2022role}, swirling flows \citep{moczarski2022interaction}, bluff body wakes \citep{nekkanti_nidhan_schmidt_sarkar_2023}, forced jets \citep{maia2021nonlinear, nekkanti2022triadic,nekkanti2023bispectral}, and twin rectangular jets \citep{yeung2025spectral}.

This paper is organized as follows. In \S \ref{sec:Methods_C-BMD}, the numerical methodology for BMD is discussed and the symmetries of the bispectrum are examined. Results from BMD analysis are presented in section \S \ref{sec:results}. Section \S \ref{sec:spatial_region} details the spatial localization of energy transfer for different triads. The results are summarized in \S \ref{sec:conclusions}.

\section{Cross-bispectral mode decomposition} \label{sec:Methods_C-BMD}

We employ the cross-bispectral mode decomposition (C-BMD), recently developed by \cite{schmidt2020bispectral}, to identify the spatially coherent structures arising from triadic interactions in a turbulent flow. Here, we provide a brief overview of the method. The reader is referred to \citet{schmidt2020bispectral} for further details of the derivation and mathematical properties of the method.


Given a statistically stationary flow field, let $\vb{q}_i =\vb{q}(t_i)$ denote the mean subtracted snapshots, where $i =1,2,\cdots n_t$ are $n_t$ number of snapshots. For spectral estimation, using Welch's approach, the dataset is first segmented into $n_{\rm{blk}}$ overlapping (by 50\%) blocks with $n_{\rm{fft}}$ snapshots in each block. Each block then Fourier transformed in time and all Fourier realizations of the $l$-th frequency, $\vb{q}^{(j)}_l$, are arranged in a matrix,

\begin{equation}
\Hat{\vb{Q}}_{l}=\bqty{\Hat{\vb{q}}_{l}^{(1)}, \Hat{\vb{q}}_{l}^{(2)}, \cdots, \Hat{\vb{q}}_{l}^{(n_{\rm blk})} }.
\label{eq:fourier realizatin}
\end{equation}
The auto-bispectral matrix is then computed as 
\begin{equation}
    \vb{B} = \frac{1}{n_{\rm{blk}}}\hat{\vb{Q}}_{k\circ l}^{H}\vb{W}\hat{\vb{Q}}_{k+l},
\end{equation}
where $\hat{\vb{Q}}_{k\circ l}^{H} = \hat{\vb{Q}}_{k}^{*}\circ \hat{\vb{Q}}_{l}^{*}$ and $\vb{W}$ is the diagonal matrix containing the spatial quadrature weights. The auto-bispectral density matrix measures the interactions between different frequencies at the same azimuthal wavenumber. To estimate the interactions between the azimuthal wavenumber triad, [$m_1$, $m_2$, $m_3$], where $m_1 + m_2=m_3$, we construct the cross-bispectral matrix
\begin{equation}
    \vb{B}_c = \frac{1}{n_{\rm{blk}}}\pqty{\hat{\vb{Q}}_{k}^{*}\circ\hat{\vb{R}}_{l}^{*}} \vb{W}\hat{\vb{S}}_{k+l}.
\end{equation}
Here, $\hat{\vb{Q}}_{k}$, $\hat{\vb{R}}_{l}$, $\hat{\vb{S}}_{k+l}$, comprises of all the Fourier realizations at the $k$-th frequency of the azimuthal wavenumber $m_1$,  the $l$-th frequency of the azimuthal wavenumber $m_2$, and the $(k+l)$-th frequency of the azimuthal wavenumber $m_3$, respectively. Owing to the non-Hermitian nature of the bispectral matrix, the optimal expansion coefficients, $\vb{a}_1$ are obtained by maximizing the absolute value of the Rayleigh quotient of $\vb{B}_c$
\begin{equation}
    \vb{a}_1 = \arg\max\limits_{\|\vb{a}\|=1} \bigg |\frac{\vb{a}^*\vb{B}_c\vb{a}}{\vb{a}^*\vb{a}}\bigg|.
\end{equation}
The complex mode bispectrum is then obtained as  
\begin{equation}
    \lambda_1(f_k,f_l) = \bigg |\frac{\vb{a}_1^*\vb{B}_c\vb{a}_1} {\vb{a}_1^*\vb{a}_1}\bigg|.
\end{equation}
Finally, the leading-order bispectral modes and the cross-frequency fields are recovered as 
\begin{align}
    \vb*{\phi}_{k+l}^{(1)} &= \hat{\vb{S}}_{k+l}\vb{a}_{1}, \quad \text{and} \\
\vb*{\phi}_{k\circ l}^{(1)} &= \pqty{\hat{\vb{Q}}_{k} \circ \hat{\vb{R}}_{l}}\vb{a}_{1},
\end{align}
respectively.  By construction, the bispectral modes and cross-frequency fields have the same set of expansion coefficients. This explicitly ensures phase coupling between the resonant frequency triad, ($f_k$, $f_l$, $f_k + f_l$), where $\hat{\vb{Q}}_{k} \circ \hat{\vb{R}}_{l}$ is the cause and $\hat{\vb{S}}_{k+l}$ is the effect. The complex mode bispectrum, $\lambda_1$, measures the intensity of the triadic interaction, and the bispectral mode, $\vb*{\phi}_{k+l}$, represents the structures that result from the nonlinear triadic interaction. Another useful quantity is the interaction map, defined as
\begin{equation}
        \vb*{\psi}_{k,l} = |\vb*{\phi}_{k\circ l} \circ \vb*{\phi}_{k+l}|. 
\end{equation}
This interaction map identifies the spatial regions of activity of the triadic interaction for the frequency triad, $\pqty{f_k,f_l,f_{k+l}}$ and wavenumber $\bqty{m_1,m_2,m_3}$. In what follows, we differentiate between the frequency and azimuthal wavenumber triads using the notation $(\cdot,\cdot,\cdot)$ and $[\cdot,\cdot,\cdot]$, respectively. 

\subsection{Symmetries of the bispectrum}\label{sec:symmetries_bmd}

Consider the pointwise cross-bispectrum for a frequency triad, $(f_1,f_2,f_3)$ and azimuthal wavenumber triad $[m_1,m_2,m_3]$ as   
\begin{equation}
 B(f_1,f_2,m_1,m_2) = \big|E\bqty{\hat{\vb{q}}^{*}(f_1,m_1) \circ \hat{\vb{q}}^{*}(f_2,m_2) \circ \hat{\vb{q}}(f_3,m_3)}\big|
\end{equation}
where, $f_3 = f_1 + f_2$, and $m_3 = m_1 + m_2$. The bispectral modes maximize the spatial integral of the pointwise bispectrum, and share the same symmetries.  For simplicity, we suppress the azimuthal wavenumber triad notation and consider the cross-bispectrum for three different signals, $\vb{q},\vb{r},\vb{s},$ as 
\begin{equation}
 B_{\vb{q}\vb{r}\vb{s}}(f_1,f_2) = \big|E\bqty{\hat{\vb{q}}^{*}(f_1) \circ \hat{\vb{r}}^{*}(f_2) \circ \hat{\vb{s}}(f_3)}\big|.
\end{equation}
There exists different possibilities for $\vb{q},\vb{r},\vb{s}$ in a turbulent flow that is homogeneous in one or more directions, and their symmetries are discussed below. The symmetry relations between positive and negative wavenumbers and frequencies are discussed in Appendix \S \ref{appendix:1}.

\subsubsection{Case 1: $\vb{q}=\vb{r}=\vb{s} \in \mathbb{R}$}
The bispectrum is symmetric about the line $f_1 = f_2$, as  $B_{\vb{q}\vb{q}\vb{q}}(f_1,f_2) = B_{\vb{q}\vb{q}\vb{q}}(f_2,f_1)$. For a real signal, the Fourier transform satisfies $\hat{\vb{q}}(f)=\hat{\vb{q}}^{*}(-f)$, which results in the bispectrum being symmetric about the line $f_1=-f_2$: 
\begin{align*}
 B_{\vb{q}\vb{q}\vb{q}}(-f_2,-f_1) &= \big|E\bqty{\hat{\vb{q}}^{*}(-f_2) \circ \hat{\vb{q}}^{*}(-f_1) \circ \hat{\vb{q}}(-f_3)}\big| \\
&= \big|E\bqty{\hat{\vb{q}}(f_2) \circ \hat{\vb{q}}(f_1) \circ \hat{\vb{q}}^{*}(f_3)} \big| \\
& = B_{\vb{q}\vb{q}\vb{q}}^{*}(f_1,f_2). 
\end{align*}
Additionally, it can be shown that:
\begin{align}
 B_{\vb{q}\vb{q}\vb{q}}(f_1,f_2) &=  B_{\vb{q}\vb{q}\vb{q}}(f_1,-f_1-f_2) = B_{\vb{q}\vb{q}\vb{q}}(-f_1-f_2,f_1) \\
 &= B_{\vb{q}\vb{q}\vb{q}}(f_2,-f_1-f_2) = B_{\vb{q}\vb{q}\vb{q}}(-f_1-f_2,f_2).
\end{align}
Due to these symmetries, the triangle enclosed by the vertices ($f_1$,$f_2$)=$(0,0)$, $(0,f_N)$, $(f_N/2,f_N/2)$ for $f_1 \geq 0$, $f_2 \geq 0$, suffices to give us the total information. Here, $f_N$ is the Nyquist frequency. Physically, this case corresponds to the azimuthal wavenumber triad $[m_1,m_2,m_3]=[0,0,0]$.


\subsubsection{Case 2: $\vb{q}=\vb{r}=\vb{s} \in \mathbb{C}$}
Similar to the case above, the bispectrum is symmetric about the line $f_1 = f_2$ as  $B_{\vb{q}\vb{q}\vb{q}}(f_1,f_2) = B_{\vb{q}\vb{q}\vb{q}}(f_2,f_1)$. For a complex-valued signal, $\hat{\vb{q}}(f) \neq \hat{\vb{q}}^{*}(-f)$, therefore the bispectrum is not symmetric about the line $f_1=-f_2$:
\begin{align}
 B_{\vb{q}\vb{q}\vb{q}}(-f_2,-f_1) & = \big|E\bqty{\hat{\vb{q}}^{*}(-f_2) \circ \hat{\vb{q}}^{*}(-f_1) \circ \hat{\vb{q}}(-f_3)} \big| \nonumber \\
  & \neq  \big|E\bqty{\hat{\vb{q}}(f_2) \circ \hat{\vb{q}}(f_1) \circ \hat{\vb{q}}^{*}(f_3)} \big| \nonumber  \\
   & \neq B_{\vb{q}\vb{q}\vb{q}}^*(f_1,f_2) 
    \label{eq:complex_signal_same2} 
\end{align}
This case is generally not encountered in turbulent flows with different wavenumber triads, as $m_3 = m_1 + m_2$. A  commonly encountered triad is $[m,m,2m]$, i.e., $\vb{q} = \vb{r} \neq \vb{s} \in \mathbb{C}$, which is also symmetric only about the line $f_1 = f_2$. 

\subsubsection{Case 3: Three different real signals}
For three different real signals, i.e.,  $\vb{q} \neq  \vb{r} \neq \vb{s}  \in \mathbb{R}$, the cross-bispectrum exhibits $180^\circ$ rotational symmetry. Consequently, the first and third quadrants, as well as the second and fourth, contain the same information, making it sufficient to analyze only one half-plane.
\begin{align}
B_{\vb{q}\vb{r}\vb{s}}(f_1,f_2) &= \big|E\bqty{\hat{\vb{q}}^{*}(f_1) \circ \hat{\vb{r}}^*(f_2) \circ \hat{\vb{s}}(f_3)}\big|,  \nonumber\\
 B_{\vb{q}\vb{r}\vb{s}}(-f_1,-f_2) &= \big|E\bqty{\hat{\vb{q}}^{*}(-f_1) \circ \hat{\vb{r}}^*(-f_2) \circ \hat{\vb{s}}(-f_3)}\big| \nonumber\\
&= \big|E\bqty{\hat{\vb{q}}(f_1) \circ \hat{\vb{r}}(f_2) \circ \hat{\vb{s}}^*(f_3)}\big| \nonumber\\
 &= B_{\vb{q}\vb{r}\vb{s}}^*(f_1,f_2) 
\end{align}



\subsubsection{Case 4: $\vb{q} = \vb{r}^* \in  \mathbb{C}, \vb{s} \in \mathbb{R}$}
Now, let’s consider the case where $\vb{r} = \vb{q}^{*} \in \mathbb{C}$ and $\vb{s} \in \mathbb{R} $. This forms a special triad,  $[m,-m,0]$,  which contributes to the mean flow deformation. Here, $\hat{\vb{r}}(f) = \hat{\vb{q}}^{*}(-f)$, as a result, the cross-bispectrum is symmetric about the line $f_1 = -f_2$.
\begin{align}
B_{\vb{q}\vb{r}\vb{s}}(f_1,f_2) &= \big|E\bqty{\hat{\vb{q}}^{*}(f_1) \circ \hat{\vb{r}}^*(f_2) \circ \hat{\vb{s}}(f_3)}\big|,  \nonumber \\
 B_{\vb{q}\vb{r}\vb{s}}(-f_2,-f_1) &= \big|E\bqty{\hat{\vb{q}}^{*}(-f_2) \circ \hat{\vb{r}}^*(-f_1) \circ \hat{\vb{s}}(-f_3)} \big| \nonumber\\
&= \big|E\bqty{\hat{\vb{r}}(f_2) \circ \hat{\vb{q}}(f_1) \circ \hat{\vb{s}}_3^*(f_3)}\big| \nonumber\\
 &= B_{\vb{q}\vb{r}\vb{s}}^*(f_1,f_2). 
 \end{align}
 An additional symmetry arises if $\vb{q}$  and $\vb{s}$ are jointly strict-sense stationary, in which case the cross-bispectrum is symmetric about $f_1=f_2$.
\begin{align}
B_{\vb{q}\vb{r}\vb{s}}(f_1,f_2) &= \big|E\bqty{\hat{\vb{q}}^{*}(f_1) \circ \hat{\vb{r}}^*(f_2) \circ \hat{\vb{s}}(f_3)}\big| \nonumber \\ 
&= \big| E\bqty{\hat{\vb{q}}^{*}(f_1) \circ \hat{\vb{q}}(-f_2) \circ \hat{\vb{s}}(f_3)}\big| \label{eq:ccr1}  \\
B_{\vb{q}\vb{r}\vb{s}}(f_2,f_1) &= \big|E\bqty{\hat{\vb{q}}^{*}(f_2) \circ \hat{\vb{r}}^*(f_1) \circ \hat{\vb{s}}(f_3)}\big| \nonumber\\ 
&= \big|E\bqty{\hat{\vb{q}}(-f_1) \circ \hat{\vb{q}}^*(f_2) \circ \hat{\vb{s}}(f_3)}\big| \label{eq:ccr2}\\
&=  B_{\vb{q}\vb{r}\vb{s}}(f_1,f_2) \label{eq:ccr}
\end{align}
The joint strict-sense stationary property ensures that the right-hand sides of equations (\ref{eq:ccr1}) and (\ref{eq:ccr2}) are equivalent,  confirming the equality in equation (\ref{eq:ccr}).

\subsubsection{Case 5: $\vb{q} \in  \mathbb{R}, \vb{r}=\vb{s} \in \mathbb{C}$ }
A $[0,m,m]$ triad is representative of the case with $\vb{q} \in \mathbb{R}$ and $\vb{r}= \vb{s} \in \mathbb{C} $.  The cross-bispectrum exhibits the following symmetry
\begin{align}
B_{\vb{q}\vb{r}\vb{r}}(f_1,f_2) &= \big|E\bqty{\hat{\vb{q}}^{*}(f_1) \circ \hat{\vb{r}}^*(f_2) \circ \hat{\vb{r}}(f_3)}\big|,  \nonumber\\
 B_{\vb{q}\vb{r}\vb{r}}(-f_1,f_3) &= \big|E\bqty{\hat{\vb{q}}^{*}(-f_1) \circ \hat{\vb{r}}^*(f_3) \circ \hat{\vb{r}}(f_2)}\big| \nonumber\\
&= \big|E\bqty{\hat{\vb{q}}(f_1) \circ \hat{\vb{r}}(f_2) \circ \hat{\vb{r}}^*(f_3)}\big| \nonumber\\
 &= B_{\vb{q}\vb{r}\vb{r}}^*(f_1,f_2) \label{eq:shift_reflect} . 
\end{align}

This symmetry can be understood as translating the ordinate by the value of the abscissa and reflecting it across the $y$-axis.


\subsubsection{Case 6: Three different complex signals}
For three different complex signals, $\vb{q}_1 \neq  \vb{q}_2 \neq \vb{q}_3 \neq \vb{q} \in \mathbb{C}$, the cross-bispectrum does not exhibit any symmetries. 

\subsubsection{Additional symmetries}
There are additional symmetries when considering the relationship between bispectra with different azimuthal triads. The most obvious one is   
\begin{equation}
    B_{\vb{q}\vb{r}\vb{s}}(f_1,f_2) = B_{\vb{r}\vb{q}\vb{s}}(f_2,f_1), \nonumber 
\end{equation}
which indicates that the bispectrum of an azimuthal wavenumber triad $[m_1,m_2,m_3]$ can be derived from that of $[m_2,m_1,m_3]$ by reflecting it across the line $f_1=f_2$. Secondly, 
\begin{equation}
B_{\vb{q}\vb{r}\vb{s}}(f_1,f_2) = B_{\vb{q^*}\vb{r^*}\vb{s^*}}^*(-f_1,-f_2), \nonumber
\end{equation}
implying that the bispectrum of the triad $[-m_1,-m_2,-m_3]$ is a $180^\circ$ rotated version of the bispectrum corresponding to the triad $[m_1,m_2,m_3]$. Additionally,
\begin{align}
    B_{\vb{q}\vb{r}\vb{s}}(f_1,f_2) &= B_{\vb{s}\vb{r^*}\vb{q}}^*(f_1+f_2,-f_2) \nonumber \\
    &= B_{\vb{s}\vb{q^*}\vb{r}}^*(f_1+f_2,-f_1), \nonumber 
\end{align}
revealing that the bispectra of the triads $[m_3,-m_2,m_1]$ and $[m_3,-m_1,m_2]$  can be deduced from the bispectrum of $[m_1,m_2,m_3]$.  In summary, if the bispectrum of the azimuthal wavenumber triad $[1,2,3]$ is known, bispectra of the triads $[2,1,3]$, $[-1,-2,-3]$, $[-2,-1,-3]$, $[3,-1,2]$, $[3,-2,1]$, $[1,-3,2]$, and $[2,-3,-1]$ can be readily determined.  





\section{Globally dominant triads and their flow structures } \label{sec:results}

\subsection{Large-eddy simulation data} \label{sec:les}

We utilize large-eddy simulation (LES) data for an isothermal subsonic turbulent
jet, at a Mach number $M_j=U_j/c_{\infty}=0.4$ and a Reynolds number $Re=\rho_j U_j D/\mu_j = 450,000$ computed by \citet{bres2019modelling}. Here, $\rho$ is the density, $U$ velocity, $D$ nozzle diameter, $\mu$ dynamic viscosity and $c$ speed of sound. The simulations were carried out using the compressible flow solver Charles on an unstructured grid using a finite-volume method with sub-grid turbulence generation in the nozzle region \cite{bres2017unstructured,bres2018importance}. The LES database consists of 20000 snapshots sampled at an interval of $\Delta tc_{\infty}/D = 0.2$ acoustic time units. Data interpolated on a cylindrical grid spanning $x,r \in [0, 30] \times [0, 6]$ was used in this analysis. The flow is non-dimensionalized by the nozzle exit values, namely velocity by $U_j$, pressure by $\rho_j U_j^2$, length by the nozzle diameter $D$, and time by $D/U_j$. Frequencies are reported in terms of the Strouhal number $St = fD/U_j$.  We focus our analysis to the most energetic wavenumbers, $m = 0,1,2,3,4$, and frequencies,  $\St_3 \in [-2,2]$, within the domain $x/D \in [0, 20]$ and $r/D \in [0, 3]$. The higher wavenumbers and frequencies are important near the nozzle's exit and will be considered wherever relevant. 

\subsection{Nonlinear transfer and BMD} \label{sec:newsec}

Quadratic nonlinearities arise from the convective term in the Navier-Stokes equation. Corresponding, they contribute to the turbulent kinetic energy through the nonlinear transport term \citep{pope2001turbulent} given by: 
\begin{equation}
    \boldsymbol{\mathcal{T}}_{nl} = -\frac{1}{2}\frac{\partial}{\partial \vb{x}_i} \overline{\vb{u}^{\prime}_i\vb{u}^{\prime}_j\vb{u}^{\prime}_i}. 
    \label{eq:total_tnl}
    \end{equation} 
Owing to the low Mach number, we have used the incompressible form of nonlinear transport. Using the Fourier transform, the transfer for an azimuthal wavenumber triad, $[m_1,m_2,m_3]$, is given as
 \begin{equation}
    \boldsymbol{\mathcal{T}}_{nl}^{[m_1,m_2,m_3]} = -\mathcal{R}\Bigg[\overline{\hat{\vb{u}}^*_j(m_3)
     \hat{\vb{u}}_i(m_1)\frac{\partial \hat{\vb{u}}_j}{\partial \vb{x}_i}(m_2)} \Bigg].
     \label{eq:tnl_azimt}
    \end{equation}


\begin{figure}
\centering
{\includegraphics[trim={0.0cm 3.7cm 0cm 3.1cm},clip,width=1.0\textwidth]{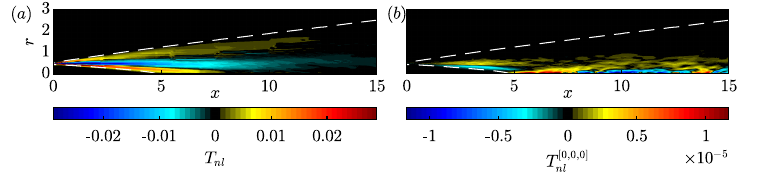}}
\caption{Spatial field of the nonlinear energy transfer: ($a$) total; ($b$) azimuthal wavenumber triad, $[m_1,m_2,m_3]=[0,0,0]$.}
\label{fig:Tnl_fields}
\end{figure}

Figure \ref{fig:Tnl_fields} shows the spatial field for the total nonlinear transfer and the wavenumber triad $[0,0,0]$. Positive (red color) and negative value (blue color) represent energy gained and lost by the fluctuations, respectively. The total nonlinear energy transfer, $\mathcal{T}_{nl}$, reveals that all nonlinear interactions occur in the shear layer and are concentrated around the lip line ($r=0.5$) within the first five jet diameters. Along $r=0.5$, the fluctuating velocities lose energy, while they gain energy in the radially outward part of the shear layer. On the other hand, for the self-interaction of the axisymmetric component, $[0,0,0]$, the nonlinearity is concentrated near the centerline beyond the end of the potential core.  Later, in section \S  \ref{sec:spatial_region}, we will demonstrate that the nonlinear activity near the nozzle's exit is due to higher wavenumber interactions, and the nonlinear activity at the end of the potential core results from interactions between lower azimuthal wavenumber.

\begin{figure}
\centering
{\includegraphics[trim={0.0cm 2.5cm 0cm 1.5cm},clip,width=1.0\textwidth]{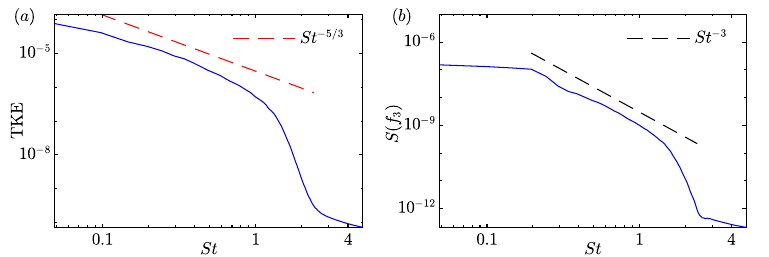}}
\caption{Turbulent kinetic energy spectrum ($a$) and bispectrum ($b$) at $x=20,r=0.5$. -5/3 and -3 scaling are shown in ($a$) and ($b$), respectively.}
\label{fig:TKE_bispectrum_scaling}
\end{figure}

Figure \ref{fig:TKE_bispectrum_scaling} shows the scaling of the turbulent kinetic energy spectrum and bispectrum in the fully developed region of the jet, i.e.,  we specifically choose the point $x=20,r=0.5$. The TKE spectrum exhibits a $St^{-5/3}$ scaling, which is characteristic of the inertial subrange of the turbulent flow. The summed bispectrum in figure \ref{fig:TKE_bispectrum_scaling}($b$) is computed as $ S(f_3) = \sum\limits_{f_1+f_2=f_3}E\bqty{\hat{\vb{u}}^{*}_i(x,f_1) \circ \hat{\vb{u}}^{*}_i(x,f_2) \circ \hat{\vb{u}}_i(x,f_3)}$. The bispectrum demonstrates a -3 scaling. This scaling is typical for the inertial range as shown by \citet{herring1980theoretical}  and \citet{van1979inertial} for isotropic turbulence in the inertial range. Later in figure \ref{fig:bispectrum_m112_x}, we will show how the pattern of the bispectrum changes from the initial shear layer to the fully developed region of the turbulent jet.

\begin{figure}
\centering
{\includegraphics[trim={0.0cm 0.4cm 0cm 0.2cm},clip,width=1.0\textwidth]{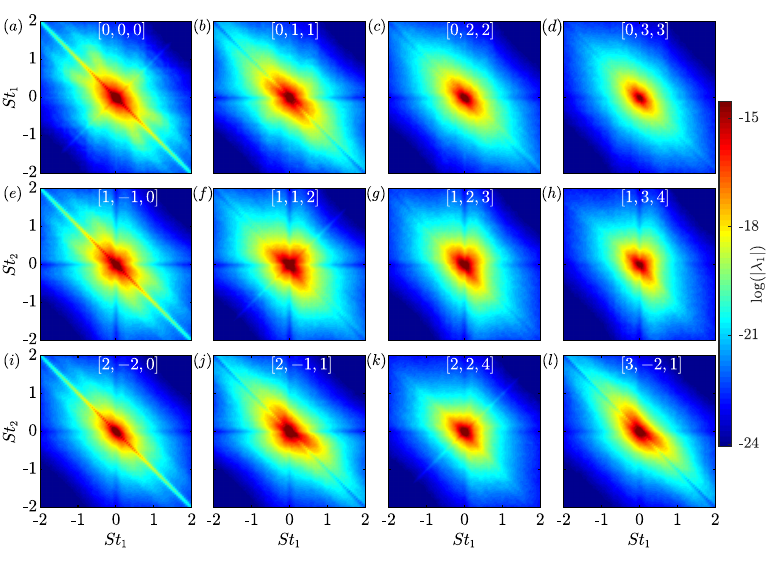}}
\caption{Cross-bispectra of twelve azimuthal wavenumber triads, $ [m_1, m_2, m_3]$. Here the magnitude of the complex cross-bispectral measure, $|\lambda_1|$, is presented.}
\label{fig:Spectra}
\end{figure}

We now perform BMD to identify the dominant triadic interactions. BMD is computed for blocks containing $n_{\rm{fft}}=256$ snapshots with $50\%$ overlap, resulting in a total number of $n_{\rm{blk}} = 158$ blocks, which represents a reasonable tradeoff between spectral resolution and converged statistics \citep{schmidt2020guide,nekkanti2023gappy}. The cross-bispectra for the twelve most dominant azimuthal triads are shown in figure \ref{fig:Spectra}. The high-intensity regions signify the dominant triads that arise from the interactions of the two displayed frequencies. All twelve cases exhibit broadband behavior, with the most intense interactions occurring at low frequencies. Similarly, in the wavenumber space, the intensity decreases for higher azimuthal wavenumber interactions, as seen in $[0,3,3]$, $[1,3,4]$, and $[2,2,4]$. A band is observed along $St_3=0$ for $[0,0,0]$, $[1,-1,0]$, and $[2,-2,0]$ triads, which is due to spectral leakage, and caution is advised while making interpretations. The cross-bispectrum for the azimuthal triads, $[0,0,0]$, $[1,-1,0]$ and $[2,-2,0]$ exhibit a four-fold symmetry about the lines, $St_1=St_2$ and $St_1=-St_2$, whereas the azimuthal triads $[1,1,2]$ and $[2,2,4]$ are symmetric about the line $St_1=St_2$. Following discussion in \S \ref{sec:symmetries_bmd}, the bispectrum of the triads, $[0,1,1]$, $[0,2,2]$ $[0,3,3]$ exhibit the reflect-translation symmetric (equation \ref{eq:shift_reflect}). The bispectra of other triads do not exhibit any symmetries. 

\begin{figure}
\centering
{\includegraphics[trim={0.0cm 2.1cm 0cm 1.2cm},clip,width=1.0\textwidth]{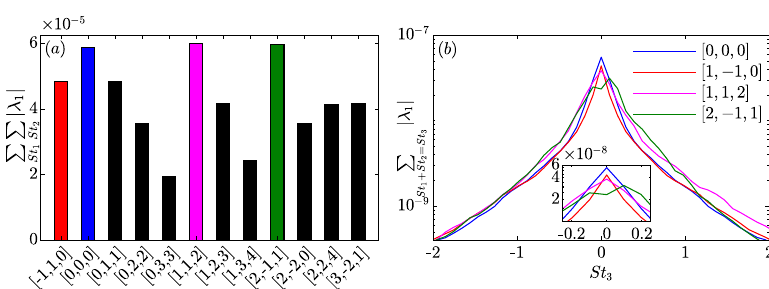}}
\caption{Cross-bispectra summed by magnitude: ($a$) over all frequency triads; ($b$) over diagonals of slope -1 such that $St_1 + St_2 =St_3$.} 
\label{fig:Spectra_Sum}
\end{figure}

Figure \ref{fig:Spectra_Sum} shows the cross-bispectrum summed over all frequency triads and summed over all diagonals of slope -1 such that $St_1 + St_2 =St_3$. The former represents the total intensity of triadic interactions for a single azimuthal wavenumber triad, while the latter denotes the intensity of the resulting frequency for a single wavenumber triad. Figure \ref{fig:Spectra_Sum}($a$) reveals that the four dominant azimuthal wavenumber triads are  $[1,1,2]$, $[0,0,0]$, $[2,-1,1]$, and $[-1,1,0]$. The self-interaction of $m=1$ that results in $m=2$ is the most significant triad. This is expected since $m=1$ and $m=2$ are the most energetic components \citep{SchmidtJFM2018}. The total intensity of the triadic interaction $[1,-1,0]$ equals  $[0,1,1]$, $[2,-2,0]$ equals $[0,2,2]$, $[1,2,3]$ equals $[3,-2,1]$, confirming the bispectrum symmetries discussed in section \ref{sec:symmetries_bmd}. The cross-bispectra, summed along diagonals, are shown in figure \ref{fig:Spectra_Sum}($b$) for the four dominant wavenumber triads. The four curves decrease monotonically. The triads $[0,0,0]$, $[1,-1,0]$, $[1,1,2]$ peak at $\St_3 \rightarrow 0$, whereas $[2,-1,1]$ peaks at the first non-zero frequency, $\St_3 \approx 0.05$, as shown in the inset of figure \ref{fig:Spectra_Sum}($b$). For $\St_3 \geq 1$, the triad $[1,1,2]$ exhibits the largest values, indicating that the strongest triadic interactions lead to the azimuthal wavenumber $m=2$ within this frequency range. The summed cross-bispectrum is symmetric about $\St_3 = 0$ for  $[0,0,0]$ and $[1,-1,0]$ triads but is not symmetric for the $[1,1,2]$ and $[2,-1,1]$ triads. This follows from the symmetry of the bispectrum, i.e., if the bispectrum is symmetric about the line $\St_1=-\St_2$, then the summed bispectrum is symmetric about $\St_3 = 0$.

\begin{figure}
\centering
{\includegraphics[trim={0.0cm 5.05cm 0cm 1.55cm},clip,width=1.0\textwidth]{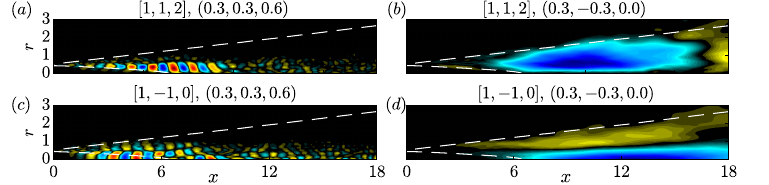}}
\caption{ Real component of the fluctuating streamwise velocity of the cross-bispectral mode for three frequency triads and three azimuthal wavenumber triads. Top ($a$,$b$)  and bottom ($c$,$d$) rows show the azimuthal wavenumber triads $[1,1,2]$, and $[1,-1,0]$, respectively. The left and right columns depict the frequency triads, $(0.3,0.3,0.6)$, and $(0.3,-0.3,0.0)$, respectively. Contours (${\color{Red}\blacksquare}\!{\color{Black}\blacksquare}\!{\color{Blue}\blacksquare}$) are given by $\pm ||\mathbf{\phi}_1: u_x ||_\infty$ of each mode.} 
\label{fig:BMD_mode}
\end{figure}

Next, we visualize the spatial flow structures associated with the dominant triads. Figure \ref{fig:BMD_mode} shows the bispectral modes due to the interaction between $m=\pm1$, $\St= \pm 0.3$. These interactions lead to four possible modes: ($i$) $[1,1,2]$, $(0.3,0.3,0.6)$; ($ii$) $[1,1,2]$, $(0.3,-0.3,0.0)$; ($iii$) $[1,-1,0]$, $(0.3,0.3,0.6)$; ($iv$) $[1,-1,0]$, $(0.3,-0.3,0.0)$. We specifically chose $m=\pm1$ and $\St= \pm 0.3$ because it exhibits the largest linear resolvent gain \citep{SchmidtJFM2018}. In both azimuthal wavenumber triads, constructive self-interference of $\St=0.3$ generates structures resembling KH wavepackets, while destructive self-interaction of $\St=0.3$ produces elongated structures localized beyond the end of the potential core. Notably, the structure in figure \ref{fig:BMD_mode}($b$) resembles a large-scale streaky structure, suggesting that the KH mode of $m=1, \St=0.3$ interacts with its conjugate $m=1, \St=-0.3$, giving rise to streaks. These observations are in line with a number of previous studies in free shear flow and jets that have postulated the creation of vortices that ultimately give rise to streaks via the interaction of KH modes \citep{pierrehumbert1982two,wu2009low}.



\begin{figure}
\centering
{\includegraphics[trim={0.0cm 2.5cm 0cm 4.6cm},clip,width=1.0\textwidth]{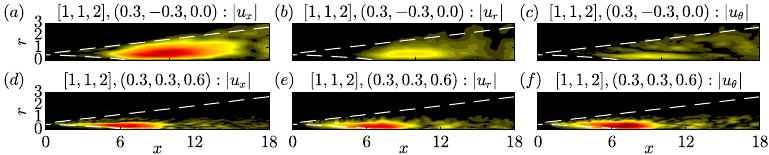}}
\caption{Magnitude of three velocity components of the cross-bispectral mode for the frequency triad, $(0.3,-0.3,0.0)$ and the wavenumber triad, [1,1,2]: ($a$) $u_x$; ($b$) $u_r$; ($c$) $u_{\theta}$.} 
\label{fig:abs_Streaks_LU}
\end{figure}

To further confirm that the structure in figure \ref{fig:BMD_mode}($b$) is a streak, the magnitude of its three velocity components are examined in figure \ref{fig:abs_Streaks_LU}. The panels in the top row reveal that the magnitude of the streamwise velocity is significantly greater (by a factor of 3) than the other two velocity components, in accordance with previous studies  \citep{boronin2013non,pickering2020lift,nekkanti_nidhan_schmidt_sarkar_2023}. For comparison, the bottom row shows the magnitude of the three velocity components for the frequency triad $(0.3,0.3,0.6)$, which corresponds to a KH-type wavepacket.  In this case, the magnitudes of the three velocity components are comparable, clearly distinguishing the characteristics of KH wavepackets from streaky structures. 

\subsection{Streaks} \label{sec:lift}

\begin{figure}
\centering
{\includegraphics[trim={0.0cm 3.6cm 0cm 1.6cm},clip,width=1.0\textwidth]{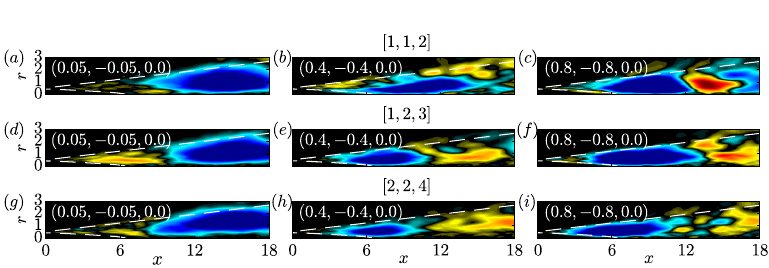}}
\caption{Real component of the fluctuating streamwise velocity of the cross-bispectral mode for three frequency triads and three azimuthal wavenumber triads. Top ($a$-$c$), middle ($d$-$f$), and bottom ($g$-$i$) rows show the azimuthal wavenumber triads [1,1,1], [1,2,3] and [2,2,4], respectively. The left, center, and right columns depict the frequency triads, $(0.05,-0.05,0.0)$, ($0.4,-0.4,0.0$), and ($0.8,-0.8,0.0$), respectively. Contours (${\color{Red}\blacksquare}\!{\color{Black}\blacksquare}\!{\color{Blue}\blacksquare}$) are given by $\pm 0.5||\mathbf{\phi}_1: u_x ||_\infty$ of each mode.}
\label{fig:Bispectral_Streaks_LU}
\end{figure}

We further investigate the role of triadic interactions in streak dynamics. Figure \ref{fig:Bispectral_Streaks_LU} shows the bispectral modes for three azimuthal wavenumber triads: $[1,1,2]$, $[1,2,3]$, and $[2,2,4]$, and three frequency triads: $(0.05,-0.05,0.0)$, $(0.4,-0.4,0.0)$, and $(0.8,-0.8,0.0)$. Streaks are azimuthally non-uniform structures and are not present in the axisymmetric component ($m = 0$). \citet{pickering2020lift} demonstrated that $m=2$, 3, and 4 are the most significant contributors to streaks. Therefore, we choose the dominant azimuthal triads that result in these wavenumbers. All nine modes display an elongated region or pronounced streamwise velocity. The lowest azimuthal wavenumber streaks are located beyond the end of the potential core at lower frequencies, shifting upstream at higher frequencies and azimuthal wavenumbers. 

\begin{figure}
\includegraphics[width=1\textwidth,trim={0cm 0.2cm 0cm 4.1cm},clip]{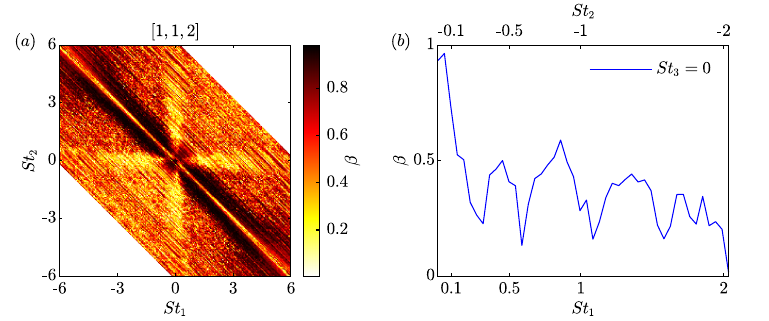}
\caption{Similarity index, $\beta$ (equation \ref{sim_index}),  of BMD modes and the leading SPOD mode interaction amongst low-frequency modes (leading SPOD mode) that result in a streak (BMD mode). Modes present the streamwise velocity reconstructed in physical space with red-blue isocontours representing $\pm0.5||\mathbf{\psi}:u_x||$ for SPOD and $\pm0.5||\mathbf{\phi}:u_x||$ for C-BMD.}
\label{fig:sim_index}
\end{figure}

The structure of the BMD streaks is qualitatively similar to those identified using SPOD.  To quantify the correspondence, we define an alignment metric between BMD modes and the leading SPOD mode as a normalized inner product, 

\begin{equation}
    \beta_{k,l} = \frac{\big|\pqty{\vb*{\phi}_{k+l}^{\text{BMD}}}^*\vb{W} \vb*{\phi}_{k+l}^{\text{SPOD}}\big|}{\sqrt{\pqty{\vb*{\phi}_{k+l}^{\text{BMD}}}^*\vb{W} \vb*{\phi}_{k+l}^{\text{BMD}}}\sqrt{\pqty{\vb*{\phi}_{k+l}^{\text{SPOD}}}^*\vb{W} \vb*{\phi}_{k+l}^{\text{SPOD}}}}.
    \label{sim_index}
\end{equation}
The similarity index, $0 \le \beta \le 1$ is unity if the structures are the same. The similarity of BMD modes of the $[1,1,2]$ triad with the leading SPOD mode of $m=2$ is shown in figure \ref{fig:sim_index}. Figure \ref{fig:sim_index}($a$) shows that the BMD modes associated with difference interactions, i.e., $(\pm\St_1, \mp\St_2)$, denoted by black color, exhibit high similarity with the leading SPOD mode. This suggests that the difference interactions, rather than sum interactions, exhibit flow structures corresponding to the most energetic coherent structures. Additionally, BMD modes generated due to self-interaction along the line $\St_1=\St_2$ for $|\St_1|=|\St_2| \leq 0.45$ exhibit high similarity, implying that self-interactions in this frequency range also result in the most energetic coherent structures. Since streaky structures are present along the line $\St_1 + \St_2=0$, the similarity index $\beta$ along this line is shown in figure \ref{fig:sim_index}($b$). The triadic interaction $[1,1,2]$, $(0.05,-0.05,0.0)$ has the highest similarity with the leading SPOD mode. This is expected as the spatial support of the frequency $St=0.05$ is present downstream where the most energetic streaky structure is also present. This suggests that, although any frequency and its conjugate can interact to produce streaky structures, the most energetic streak occurs due to the interaction of the lowest frequency and its conjugate.

\begin{figure}
\raggedright$(a)$ 
\hspace{0.2cm} 
\includegraphics[width=1\textwidth,trim={0cm 1.2cm 0cm 0.9cm},clip]{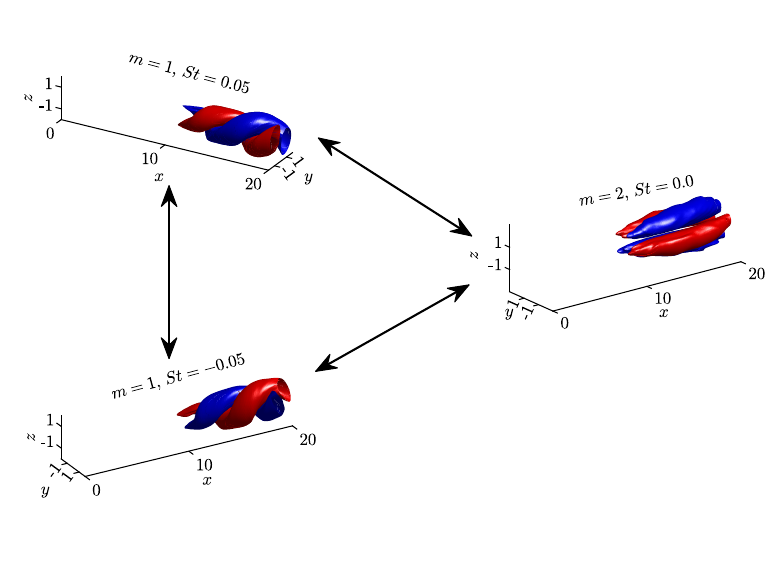}
\hspace{1cm}  
$(b)$ \hspace{0.2cm} 
\includegraphics[width=1\textwidth,trim={0cm 1.2cm 0cm 0.9cm},clip]{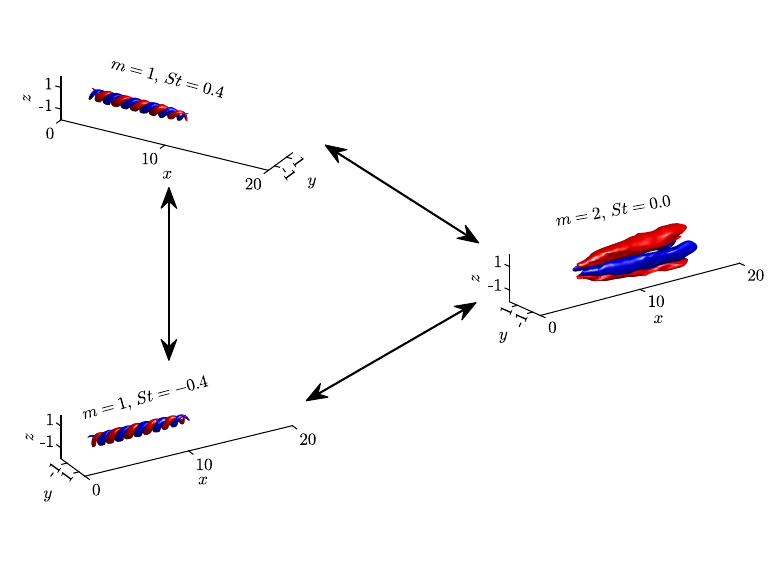}
\caption{Schematic of the triadic interaction between KH wavepacket, its conjugate, and streak for the azimuthal wavenumber triad $[m_1,m_2,m_3]=[1,1,2]$ for two frequency triads: ($a$) $(0.05,-0.05, 0.0)$; ($b$) $(0.4,-0.4, 0.0)$. Modes present the streamwise velocity reconstructed in physical space with red-blue isocontours representing $\pm0.5||\mathbf{\psi}:u_x||$ for SPOD and $\pm0.5||\mathbf{\phi}:u_x||$ for C-BMD.}
\label{fig:SCH_KH_LU}
\end{figure}

Figure \ref{fig:SCH_KH_LU} shows the schematic representation of the frequency triads $(0.05, -0.05, 0.0)$ and $(0.4, -0.4, 0.0)$ for the azimuthal wavenumber triad $[1, 1, 2]$. In panel ($a$), the structure corresponding to $m=1$, $St=\pm 0.05$ is the leading SPOD mode, while the structure corresponding to $m=2$, $St=0.0$ is the BMD mode. This schematic illustrates how two helical modes, which are complex conjugates of each other, are coupled to streaks. A similar triad, but at a higher frequency, belonging to the KH range, is shown in figure \ref{fig:SCH_KH_LU}($b$). The structure associated with $m=1$, $St=\pm 0.4$ occurs upstream, causing the streaks associated with this triadic interaction to be located further upstream compared to panel ($a$). A previous study by \citet{wu2009low} found that perturbing the flow at $m = \pm 1$ and the same frequency led to large, streamwise elongated structures that modulated the mean flow. Similar findings were reported by \citet{corke1993resonance}, \citet{kusek1990seeding}, and \citet{samimy2007active}. These studies, along with the current findings, suggest that the difference interactions between the helical structures of the same frequency can lead to streaks, which are captured by cross-bispectral modes.

\subsection{Spectral energy budget of streaks}
Now, to quantify the contribution of the linear and nonlinear terms to streaks, we use the spectral TKE equation,
\begin{equation}
 \def\?{\vphantom{\displaystyle\sum_{i=1}^N}}
    \frac{\partial \hat{k}}{\partial t}    = \mathcal{R} \bqty{  \underbrace{\?-\bar{u}_i \frac{\partial \hat{k}}{\partial x_i}}_{\mathcal{A}} \underbrace{\?- \overline{\hat{u}_j^*\widehat{ u_i\frac{\partial u_j}{\partial x_i}}}}_{\mathcal{T}_{nl}} \underbrace{\?-\overline{\hat{u}_j^*\hat{u}_i\frac{\partial \bar{u}_j}{\partial x_i}}}_{\mathcal{P}} \underbrace{\? - \frac{2}{Re} \overline{\hat{s}_{i j}^* \hat{s}_{i j}}}_{ \mathcal{D}} \underbrace{\? -\frac{\partial}{\partial x_j}\pqty{\overline{\hat{u}_j^*\hat{p}}}}_{\mathcal{T}_{p}} \underbrace{\?+\frac{2}{Re} \frac{\partial}{\partial x_i} \pqty{\overline{\hat{u}_j^*\hat{s}_{ij}}}}_{\mathcal{T}_{\nu}}}\label{eq:stke_budget}
 \end{equation}
where, $\hat{s}_{ij} = 1/2 \pqty{\partial \hat{u}_i/\partial x_j + \partial \hat{u}_j/\partial x_i }$ is the spectral strain rate, $\mathcal{R}$ denotes the real part, and $\hat{k}(f,m)= \overline{\hat{u}_i^{*}(f,m)\hat{u}_i(f,m)}/2$. For statistically stationary flows, the left-hand side is zero. On the right-hand side, the terms, $\mathcal{A}$, $\mathcal{T}_{nl}$, $\mathcal{P}$, $\mathcal{D}$, $\mathcal{T}_{p}$, and $\mathcal{T}_{\nu}$ denote the advection, nonlinear transfer, production, dissipation, pressure transport, and viscous transport, respectively. Readers are referred to \citet{bolotnov2010spectral} and \citet{nekkanti2024nonlinear} for the derivation of the spectral TKE equation.  The linear contribution to streaks, i.e., the lift-up mechanism, is through the production term 
 \begin{equation}
     \mathcal{P}\pqty{f , m} =  -\mathcal{R}\overline{\bqty{\hat{u}_j^*\pqty{f,m}\hat{u}_i\pqty{f,m}\frac{\partial \bar{u}_j}{\partial x_i}}},
     \label{P_term}
 \end{equation}
and the triadic contribution is through the nonlinear transfer term 
 \begin{equation}
     \mathcal{T}_{nl}\pqty{f, m} =  - \mathcal{R}\overline{\bqty{\hat{u}_j^*\pqty{f , m}\widehat{u_i\frac{\partial u_j}{\partial x_i}}\pqty{f, m}}},
     \label{Tnl_term1}
 \end{equation}
where $\widehat{u_i\frac{\partial u_j}{\partial x_i}}$ contains the interactions between different frequencies and azimuthal wavenumbers. Following \citet{cho2018scale}, we isolate the energy transfer of individual triads, i.e., frequency and wavenumber triplets related by the resonance conditions, $m_1 \pm m_2 \pm m_3=0$ and $f_1 \pm f_2  \pm f_3=0$ by splitting  $\widehat{u_i\frac{\partial u_j}{\partial x_i}}$ using the discrete convolution to obtain
 \begin{equation}
     \mathcal{T}_{nl}\pqty{f_3 , m_3} =  - \mathcal{R}\overline{\bqty{\hat{u}_j^*\pqty{f_3,m_3}\sum\limits_{\substack{f_1 +f_2=f_3 \\ m_1 +m_2=m_3}}\hat{u}_i\pqty{f_1,m_1}\frac{\partial \hat{u}_j}{\partial x_i}\pqty{f_2,m_2}}}.
     \label{Tnl_term2}
 \end{equation}

  \begin{equation}
     \mathcal{T}_{nl}\pqty{St=0 , m=2} =  - \mathcal{R}{\bqty{\overline{\sum\limits_{St_1,m_1}\hat{u}_j^*\pqty{St=0,m=2}\hat{u}_i\pqty{St_1,m_1}\frac{\partial \hat{u}_j}{\partial x_i}\pqty{-St_1,2-m_1}}}}.
     \label{Tnl_term2}
 \end{equation}
 
\begin{figure}
\centering
{\includegraphics[trim={0.0cm 4.55cm 0cm 3.4cm},clip,width=1.0\textwidth]{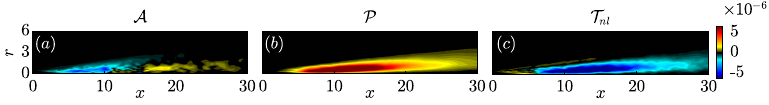}}
\caption{Spectral TKE budget, equation (\ref{eq:stke_budget}) for $m=2$, $\St=0$: ($a$) Advection, $\mathcal{A}$; ($b$) production, $\mathcal{P}$; ($c$) nonlinear transfer, $\mathcal{T}_{nl}$. } 
\label{fig:tke_budget}
\end{figure}

The incompressible turbulent kinetic energy budget is performed for the azimuthal wavenumber $m=2$ and frequency $\St=0$. Only advection, production, and nonlinear transfer are significant, and we confirmed that the sum of all terms on the right-hand side of the equation (\ref{eq:stke_budget}) is likewise negligible. Figure \ref{fig:tke_budget} shows the three relevant terms, of which $\mathcal{P}$ and $\mathcal{T}_{nl}$ are more dominant. Interestingly, these two fields are spatially similar but have opposite signs. This indicates that production supplies energy to streaks, while nonlinear transfer extracts energy from them.  The production term represents the lift-up mechanism, in which streamwise vortices (or rolls, as identified from the forcing modes in resolvent analysis \citep{pickering2020lift}) lift up high-speed fluid from the jet center and bring down low-speed fluid from the outer jet, inducing streaks.  The cumulative effect of nonlinear interactions is the backscatter of energy from the streaks, which ultimately leads to their breakdown. Our results corroborate the concept of the self-sustaining cycle \citep{hamilton1995regeneration, waleffe1992nature, brandt2014lift}, where the linear lift-up process supplies energy to the streaks and the nonlinear branch draws energy from them. Specifically, it has been shown that the nonlinear term drains energy from streaks and feeds it back to the rolls \citep{waleffe1997self, waleffe1998streamwise, dessup2018self}.  Finally, the advection term reveals that it extracts energy in the region $3\leq x \leq10$ and supplies energy in the region $x\geq 16$. However, this term is of much lower significance than production and nonlinear transfer, serving primarily to balance the energy budget.  The findings in this section suggest the following process: (i) nonlinear interactions of KH wavepackets generate streamwise vortices; (ii) these vortices transfer energy to the streaks through the linear lift-up mechanism; and (iii) the streaks ultimately breakdown due to further nonlinear interactions that drain their energy.

\section{Spatial regions of nonlinear energy transfer and energy cascade} \label{sec:spatial_region}
We next focus on identifying the spatial regions where the dominant triadic interactions occur and investigate the direction of energy transfer in azimuthal wavenumber space. We will use the terms \emph{inverse} and \emph{forward} energy cascade to imply the energy transfer from higher azimuthal wavenumbers(or higher frequencies) to lower azimuthal wavenumbers (or lower frequencies) and vice versa.

\begin{figure}
\centering
{\includegraphics[trim={0.0cm 3.85cm 0cm 1.8cm},clip,width=1.0\textwidth]{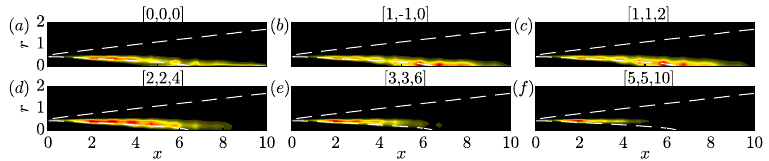}}
\caption{ Integral interaction maps, $\sum_{k,l} \lambda_1(St_k,St_l) | \phi_{k \circ l} \circ \phi_{k+l}|$, for azimuthal triads, $[m_1,m_2,m_3]$:  ($a$) $[0,0,0]$; ($b$) $[1,-1,0]$; ($c$) $[1,1,2]$;  ($d$) $[1,2,3]$; ($e$) $[2,2,4]$; ($f$) $[5,5,10]$.} 
\label{fig:hadamard_maps}
\end{figure}

We first identify the regions where different azimuthal wavenumber triads are most active. To achieve this, the weighted interaction maps are summed over all frequency triads as $\sum\limits_{k,l} \lambda_1(St_k,St_l) | \phi_{k \circ l} \circ \phi_{k+l}|$. Figure \ref{fig:hadamard_maps} shows the weighted interaction maps for six azimuthal wavenumber triads. The top row highlights three out of the four dominant wavenumber triads, while the bottom row illustrates interactions resulting in higher azimuthal wavenumbers. For the dominant triads, the interaction is present in the shear layer and beyond the end of the potential core. The region of nonlinear activity extends farthest downstream for the $[0,0,0]$ triad.  Our findings agree with the work of  \cite{tissot2017sensitivity}, demonstrating that nonlinearity becomes pronounced downstream of the potential core for the axisymmetric component.  The interaction region shifts closer to the nozzle exit for interactions involving higher azimuthal wavenumbers,  further supporting our observation from figure \ref{fig:Tnl_fields} that interactions of higher wavenumbers play a more significant role in the nonlinearities closer to the nozzle exit.

\begin{figure}
\centering
{\includegraphics[trim={0.0cm 3.8cm 0cm 1.9cm},clip,width=1.0\textwidth]{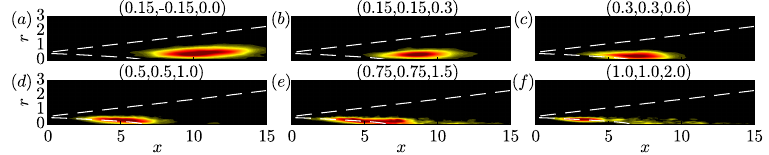}}
\caption{ Interaction maps, $| \phi_{k \circ l} \circ \phi_{k+l}|$, of the azimuthal wavenumber triad [1,1,2] for different frequency triads, $(St_1,St_2,St_3)$:  ($a$) $(0.15,-0.15,0.0)$; ($b$) $(0.15,0.15,0.0)$; ($c$) $(0.3,0.3,0.0)$;  ($d$) $(0.5,0.5,1.0)$; ($e$) $(0.75,0.75,1.5)$; ($f$) $(1.0,1.0,2.0)$.} 
\label{fig:hadamard_maps_Freq}
\end{figure}

Figure \ref{fig:hadamard_maps_Freq} displays the interaction maps for six representative frequency triads of the most dominant azimuthal wavenumber triad $[1,1,2]$. This figure highlights the regions where nonlinear interactions are most active in frequency space, complementary to the wavenumber interactions shown in figure \ref{fig:hadamard_maps}. The interaction maps indicate that significant nonlinear interaction between lower frequencies occur downstream, whereas, for interactions involving higher frequencies, this region shifts toward nozzle exit. Overall, figures \ref{fig:hadamard_maps} and \ref{fig:hadamard_maps_Freq} reveal that small-scale structures are more actively involved in nonlinear interactions in the initial shear layer, whereas large-scale structures become more active downstream. This observation is not surprising, as due to the spreading of the shear layer, structures with larger wavelengths are supported farther downstream than those associated with smaller wavelengths.

\begin{figure}
\centering
{\includegraphics[trim={0.0cm 0.0cm 0cm 0.0cm},clip,width=1.0\textwidth]{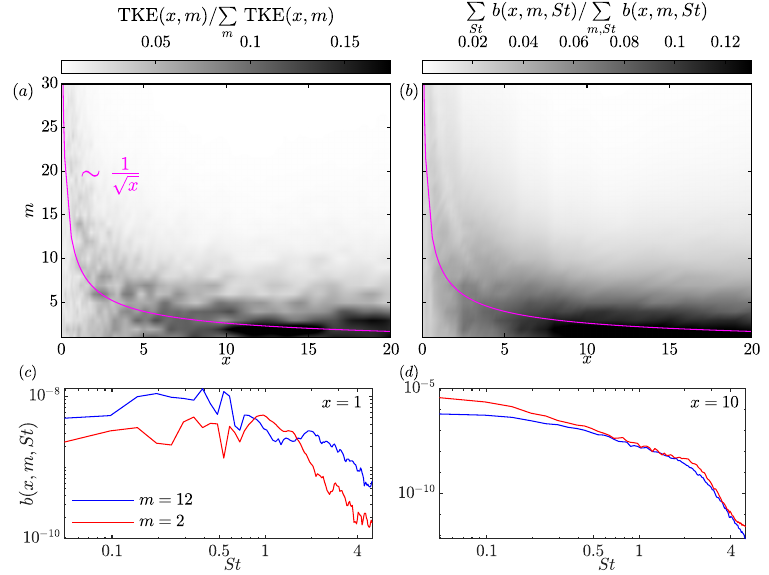}}
\caption{ Normalized TKE ($a$) and bispectrum ($b$) as a function of streamwise location and azimuthal wavenumber}
\label{fig:TKE_X}
\end{figure}

Figure \ref{fig:TKE_X} shows the normalized turbulent kinetic energy (TKE) and bispectrum as a function of streamwise location. The TKE is computed at each spatial location and then integrated radially for each azimuthal wavenumber. The bispectral measure, $b(x,m,St)$, is computed as 
\begin{equation}
    b(x,m,St) = \sum\limits_{m_1,St_1} \int\limits_r  \vb{\hat{u}}_i(x,r,m_1,St_1)\circ\vb{\hat{u}}_i(x,r,m-m_1,St-St_1)\circ \vb{\hat{u}}_i^*(x,r,m,St) rdr.
\end{equation}
 Finally, at each streamwise location, the TKE and bispectral measure are normalized by total TKE and total bispectral measure, $\sum_{m,St} b\pqty{x,m,St}$.  A similar trend is observed in both the TKE and bispectrum. Near the nozzle exit, higher azimuthal wavenumbers are more energetic and show greater triadic activity. As the flow progresses downstream, the energy and bispectral measure shift to lower azimuthal wavenumbers. The most energetic and triadically dominant azimuthal wavenumber scale as $1/\sqrt{x}$. The consistent scaling and trends between figure \ref{fig:TKE_X}($a$) and ($b$) suggest that the azimuthal wavenumbers involved in the strongest nonlinear interactions are also the most energetic. Previous studies \citep{arndt-long-glauser-pod-97, citriniti-george-pod-2000, sasaki2017high} have shown that in the region  $x\leq3$, modes up to $m \leq 10$ contain significant energy. Downstream of this region, modes with $m > 10$ lose energy while those with $m < 10$ gain energy. Eventually, in the self-similar region, most of the energy is found in $m=0$, 1, and 2. This explains why higher azimuthal wavenumbers are more involved in nonlinear interactions upstream, while lower azimuthal wavenumbers are more active downstream.

To illustrate how the bispectral measure varies across different frequencies, Figures \ref{fig:TKE_X}($c$) and ($d$) show the bispectral measure at $x=1$ and 10 for $m=2$ and 12. At $x=1$, the bispectral measure for $m=2$ and $m=12$ exhibits peaks, with global maxima at $\St=0.4$ and $\St=1$, respectively. Further downstream at $x=10$, the bispectral measure for both $m=2$ and $m=12$ becomes broadband. Since the flow is developing in the initial shear layer, the bispectral measure exhibits peaks at relatively higher frequencies. Downstream of the potential core, the flow becomes fully developed, and the bispectrum is broadband. This behavior resembles the inertial subrange energy spectrum, indicating a forward energy cascade in this region.


\begin{table*}
\caption{Dominant azimuthal wavenumber triad at different streamwise locations.}
\vspace{-1.25em}
\begin{tabular}{cccccccccccc}
\hline
\begin{tabular}[c]{@{}c@{}}Axial\\ location\end{tabular} & $x=1$ & $x=2$ & $x=3$ & $x=4$ & $x=5$ & $x=6$ & $x=8$ & $x=11$ & $x=14$ & $x=17$ & $x=20$ \\ \hline
\begin{tabular}[c]{@{}c@{}}Dominant\\ triad\end{tabular} & $[6,6,12]$ & $[3,3,6]$ & $[2,2,4]$ & $[1,1,2]$ & $[1,1,2]$ & $[1,1,2]$ & $[1,1,2]$ & $[2,2,4]$ & $[2,2,4]$ & $[2,2,4]$ & $[2,2,4]$ \\ \hline
\end{tabular}

\label{table:dom_triads}
\end{table*}

Table \ref{table:dom_triads} tracks the dominant azimuthal wavenumber triads as the flow progresses downstream. To identify these dominant triads, we first compute the spatial triple correlation of velocity components for various azimuthal wavenumber triplets. Next, we average these correlations over time. Finally, the triads with the highest values are selected. Note that for this spatial triple correlation, the intensity of the triad $[m_1, m_2, m_3]$ is equal to the intensities of $[m_3, -m_2, m_1]$, $[m_3, -m_1, m_2]$, and $[-m_1, -m_2, -m_3]$. Therefore, only the positive triads $[m_1, m_2, m_3]$ where $m_1, m_2, m_3 \geq 0$ are reported in Table \ref{table:dom_triads}. Additionally, the $[0, 0, 0]$ triad is omitted from the analysis, as it does not provide information on the direction of energy transfer in azimuthal wavenumber space. In the initial shear layer at $x=1$, the most dominant triad is $[6,6,12]$. At $x=1$, $m=12$ has the highest TKE, suggesting that the most energetic wavenumber is also involved in the dominant nonlinear interactions. Moving downstream, the dominant azimuthal triad transitions to $[3,3,6]$ at $x=2$, $[2,2,4]$ at $x=3$, and $[1,1,2]$ at $x=4$, remaining the same until $x=8$. The transition of the dominant triad from higher to lower azimuthal wavenumbers is evident up to $x \approx4$, suggesting that the inverse energy cascade is active in the developing region of the jet. Downstream of the potential core, the dominant triad increases to $[2,2,4]$, indicating that the forward cascade is active in the fully developed region. This is not surprising, as the TKE spectrum in this region is broadband and exhibits inertial subrange scaling (see figure \ref{fig:TKE_bispectrum_scaling}), which is typical of a forward energy cascade.

\begin{figure}
\centering
{\includegraphics[trim={0.0cm 1.95cm 0cm 0.9cm},clip,width=1.0\textwidth]{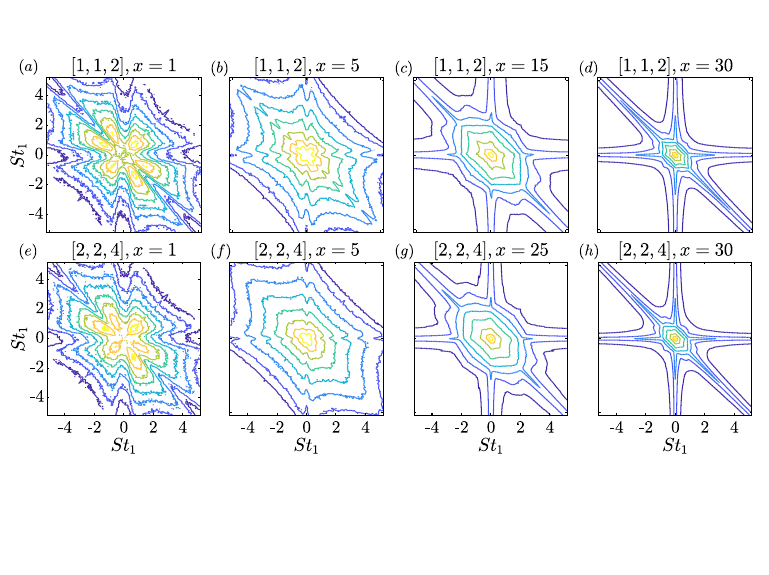}}
\caption{Mode bispectra for the azimuthal wavenumber triads, $[1,1,2]$ ($a$-$d$), and $[2,2,4]$ ($e$-$f$), at different streamwise locations: ($a$,$e$) $x=1$; ($b$,$f$) $x=5$; ($c$,$g$) $x=15$;  ($d$,$h$) $x=30$. }
\label{fig:bispectrum_m112_x}
\end{figure}

Next, we examine the patterns of the bispectrum in different spatial regions. Figure \ref{fig:bispectrum_m112_x} shows the mode bispectrum of two azimuthal wavenumber triads, $[1,1,2]$ and $[2,2,4]$, computed at four different streamwise locations: $x=1$, 5, 15, and 30, i.e., ranging from the initial shear layer to the self-similar region. At $x=1$, the bispectra in ($a$,$e$) exhibits a six-lobed pattern, with lobes along the lines $\St_1=\St_2$, $\St_1=-2 \St_2$, and $\St_1=- \St_2/2$. Previous work by \citet{herring1992spectral} on decaying turbulence also identified a similar bispectrum pattern, which they referred to as a ``six-leaf rose''. As the flow evolves downstream, the six lobes coalesce into one lobe about $(\St_1,\St_2) = (0,0)$. This is expected, as the $m=1$ mode in the initial shear layer exhibits tones. The sum interaction of these tones results in lobes along the line $\St_1=\St_2$, while the difference interactions generate lobes along the lines $\St_1=-2\St_2$ and $\St_1=- \St_2/2$. Downstream of the potential core, the spectrum of $m=1$ becomes broadband, and its self-interaction results in a single lobe at the origin.  Similar trends are observed for the $[2,2,4]$ triad. This figure implies that for a developing flow, the bispectrum pattern is similar to figure \ref{fig:bispectrum_m112_x}($a$), while for a fully developed broadband flow, it is similar to figure \ref{fig:bispectrum_m112_x}($f$).

\begin{figure}
\centering
{\includegraphics[trim={0.0cm 1.8cm 0cm 1.0cm},clip,width=1.0\textwidth]{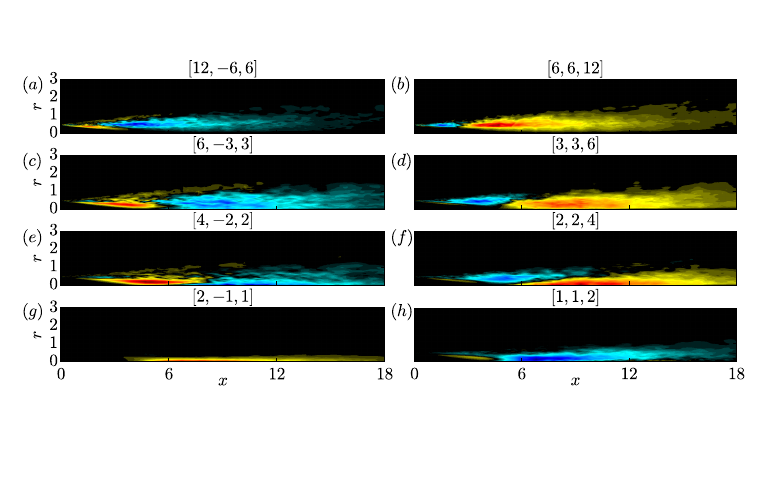}}
\caption{ Nonlinear transfer fields for different azimuthal triads, [$m_1,m_2,m_3$]: ($a$) [$12,-6,6$]; ($b$) [$6,6;12$]; ($c$) [$6,-3,3$];  ($d$) [$3,3,6$]; ($e$) [$4,-2,2$]; ($f$) [$2,2,4$]; ($g$) [$2,-1,1$]; ($h$) [$1,1,2$].}
\label{fig:Energy_transfer_Triad}
\end{figure}

Next, the direction of energy transfer for the triads reported in table \ref{table:dom_triads} is investigated. Figure \ref{fig:Energy_transfer_Triad} shows the nonlinear energy transfer, equation (\ref{eq:tnl_azimt}), for eight triads,  $[12,-6,6]$, $[6,6,12]$, $[6,-3,3]$, $[3,3,6]$, $[4,-2,2]$, $[2,2,4]$ , $[2,-1,1]$, and $[1,1,2]$. For a triad $[m_1,m_2,m_3]$, the red color denotes the energy transfer from $m_1$ and $m_2$ to $m_3$ and the blue color denotes the energy extracted from $m_3$ by $m_1$ and $m_2$. The spatial fields in the left column display a region of positive nonlinear transfer, followed by a region of negative energy transfer. In contrast, the energy transfer patterns in the right column are opposite to those in the left column. However, the panels in two columns convey the same information. For example, the spatial field corresponding to the $[12,-6,6]$ triad demonstrates energy transfer from $m= 12$ to 6 until $x \approx 2.8$,  after which there is a backscatter of energy from $m=6$ to $m=12$.  Similarly, the $[6,6,12]$ triad indicates that initially, $m=6$ extracts energy from $m=12$, but later, the energy transfers from $m=6$ to $m=12$. These observations clearly show that the inverse energy cascade is active in the initial shear layer, while the forward energy cascade becomes prominent at more downstream locations. 

Interestingly, the spatial fields in the bottom row, panels ($g$) and ($h$), demonstrate that the energy transfer is unidirectional, with $m=2$ supplying energy to $m=1$. It is noted that within the current streamwise extent ($x \leq 30$), there is no observed energy transfer from $m=1$ to $m=2$; however, this does not rule out the possibility of such transfer occurring farther downstream. Additionally, the triad $[1,-1,0]$ reveals that $m=0$ extracts energy from $m=1$, not shown here for brevity. To summarize, for the spatial domain under investigation, $m\geq2$ are involved in both forward and inverse energy cascade, while the azimuthal wavenumbers  $m=0$, and 1 act as energy sinks. Specifically, $m=1$ gains energy from $m\geq2$ and $m=0$ gains energy from all $m$.


\begin{figure}
\centering
{\includegraphics[trim={0.0cm 2.4cm 0cm 2.9cm},clip,width=1.0\textwidth]{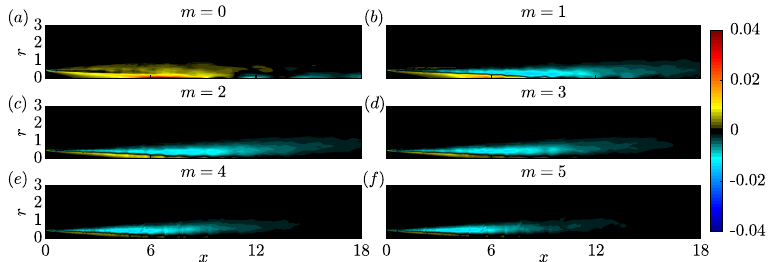}}
\caption{Net nonlinear energy transfer fields, equation (\ref{eq_netTnl}), for different azimuthal wavenumber: ($a$) $m=0$; ($b$) $m=1$; ($c$) $m=2$;  ($d$) $m=3$; ($e$)  $m=4$; ($f$) $m=5$.}
\label{fig:net_Energy_transfer_Triad}
\end{figure}

Figure \ref{fig:net_Energy_transfer_Triad} shows the net nonlinear energy transfer into the azimuthal wavenumber $m$. This is computed as 
\begin{equation}
    \mathbb{T}_{nl}(m) = \sum\limits_{m_1}  -\mathcal{R}\Bigg[\overline{\hat{\vb{u}}^*_j(m)
     \hat{\vb{u}}_i(m_1)\frac{\partial \hat{\vb{u}}_j}{\partial x_i}(m-m_1)} \Bigg]. \label{eq_netTnl}
\end{equation}
All the triadic interactions supply energy to the axisymmetric component $m=0$. For $m=1$ and $m=2$, triadic interactions extract energy in the shear layer but supply energy near the centerline towards the end of the potential core. This trend is also observed for $m=2$. For higher azimuthal wavenumbers, the loss of energy is more prominent than the gain. This is more clearly reflected in panels ($c$)-($f$). Although individual triads both supply and extract energy to each azimuthal wavenumber, as shown in figure \ref{fig:Energy_transfer_Triad}, the net triadic interactions result in an energy gain for lower azimuthal wavenumbers and an energy loss for higher azimuthal wavenumbers.

\section{Discussions and Conclusions} \label{sec:conclusions}

Triadic interactions transfer energy between scales in turbulent flows. Using bispectral mode decomposition, we characterize these interactions in a Mach 0.4 turbulent jet.  Our findings show that BMD can effectively identify the dominant triadic interactions and extract related flow structures, even in broadband turbulence.

The Kelvin-Helmholtz mechanism, the most amplified linear mechanism, is the primary component of triadic interactions. BMD reveals that its constructive self-interaction leads to higher helical structures, while its destructive self-interaction leads to elongated structures. Specifically, we identify a strong triadic interaction between a Kelvin-Helmholtz wavepacket, its complex conjugate, and streaks, spanning various azimuthal wavenumbers.  The most dominant interactions occur between $(m=1,\St)$, $(m=1, -\St)$, and $(m=2, \St=0)$. While this interaction is present across a broad frequency range, the most energetic streaks arise from lower-frequency interactions, such as the triad $(0.05,-0.05,0.0)$. These findings provide the first direct data verification of the hypothesis that vortices, ultimately giving rise to streaks, are formed through the interaction of KH modes \citep{Benney1960,widnall1974instability,pierrehumbert1982two}. This behavior is further supported by \citet{wu2009low} and \citet{zhang2023nonlinear}, who demonstrated that the quadratic interaction of helical modes generates streaks. Furthermore, a spectral energy budget analysis of streaks reveals that the contribution of nonlinear interactions is comparable in magnitude to the linear production term. While the linear mechanism supplies energy to streaks, the nonlinear interactions serve to extract energy from them. This energy loss due to the triadic interactions is analogous to the self-sustaining mechanism in wall-bounded flows, where the nonlinear interactions drain energy from streaks and feed it back to the rolls \citep{waleffe1997self, waleffe1998streamwise}.  Note that this is not inconsistent with the observation that interactions among KH wavepackets lead to streak formation. Essentially, two distinct types of nonlinear interactions are active. First, triadic interactions among KH wavepackets provide a forcing for the linear lift-up mechanism that subsequently produces streaks. Second, once the streaks have formed, nonlinear interactions become significant again, causing an energy loss that ultimately leads to their breakdown.

Next, a global and local analysis of triadic interactions across different scales is presented. Globally, across the spatial domain considered, the triadic interactions are most prominent among lower azimuthal wavenumbers $m \leq 2$, and lower frequencies $\St \leq 0.5$. This finding suggests that in a turbulent jet with a tripped boundary layer, the larger scales, which exhibit the highest energy, are also the most actively engaged in triadic interactions. In contrast, the local analysis reveals a distinct spatial variation: triadic interactions among smaller scales are more pronounced in the initial shear layer, whereas interactions among larger scales dominate downstream, beyond the potential core. More specifically, in the shear layer region from $0 \leq x \leq 6$, the dominant azimuthal wavenumber triad transitions from $[6,6,12]$ to $[3,3,6]$, then to $[2,2,4]$, and finally to $[1,1,2]$.  However, this trend reverses in the fully developed region $x \geq 6$, with the dominant triad evolving from $[1,1,2]$ to $[2,2,4]$. By computing the nonlinear transfer term, it is found that energy is transferred from higher azimuthal wavenumbers to lower wavenumbers in the shear layer, while beyond the end of the potential core, energy transfer occurs from lower to higher wavenumbers. 
These results suggest that as the flow exits the nozzle, it initially experiences an inverse energy cascade, which then transitions to a forward energy cascade as the flow progresses downstream. 

Our results leave some open questions for future research. Previous studies by \citet{wu2009low} and \citet{suponitsky2010linear} show that the quadratic interaction of $m=\pm1$ radiates low-frequency sound at angles of $60-70^\circ$ relative to the jet axis. Therefore, one possible direction is to investigate the effect of nonlinear interactions on the far-field jet noise. Another promising avenue is to examine the impact of nonlinear interactions on mean-flow distortion \citep{long1992controlled}.  This distortion is particularly pronounced in non-axisymmetrically forced jets. Investigating how such forcing mechanisms \citep{corke1991mode,long1992controlled,nekkanti2023bispectral} contribute to mean flow distortion could yield valuable insights. It is also worth investigating whether an interaction similar to KH-streak interaction occurs in supersonic jets and if nonlinear interactions exist between screech tones and streaks. 


\section*{Acknowledgements}
 The authors are grateful for support from the United States Office of Naval Research under contract N00014-23-12650 with Dr. Steven Martens as program manager. OTS acknowledges support from the ONR grant N00014-20-1-2311.
 \appendix

\section{Proof of Complex Relations} \label{appendix:1}

Owing to the axisymmetry and statistical stationarity of the turbulent jet, we perform two Fourier transforms, one in the azimuthal direction and the other in time. For a real signal $s\pqty{{\theta,t}}$, performing FFT in $\theta$ gives,
\begin{align}
    s(\theta,t) = \sum_{-m}^{m} s_m(t) e^{i m \theta},
\end{align}
where  $s_{-m}\pqty{t} = s_{m}^*\pqty{t}$, and $\pqty{\cdot}^{*}$ denotes the complex conjugate. We can rewrite the above as,
\begin{align}
    s(\theta,t) = s_0(t) + \sum_{m>0} \bigg[ \bigg( s^r_m(t)+i s^i_m(t) \bigg)  e^{i m \theta} + \bigg( s^r_m(t) - i s^i_m(t) \bigg)  e^{-i m \theta} \bigg],
\end{align}
where, $s^r_m(t)$ and $s^i_m(t)$ are  the real and imaginary components of $s_{m}\pqty{t}$. We now Fourier transform in time
\begin{align}
    s(\theta,t)  = \quad &  s_{0,0} + \sum_{\omega > 0} \bigg[s_{0,\omega} e^{i \omega t} + s^*_{0,\omega} e^{-i \omega t} \bigg] + \nonumber\\ & \sum_{\omega > 0} \sum_{m>0} \bigg[  \bigg( s^r_{m, \omega} +i s^i_{m, \omega} \bigg)  e^{i \pqty{m \theta + \omega t }} + \bigg( \pqty{s^{r}_{m, \omega}}^{*} + i \pqty{s^{i}_{m, \omega}}^{*} \bigg)  e^{i \pqty{m \theta - \omega t}} + \nonumber\\ &
      \bigg( s^r_{m, \omega} - i s^i_{m, \omega} \bigg)  e^{-i \pqty{m \theta - \omega t} } + \bigg( \pqty{s^{r}_{m, \omega}}^{*} - i \pqty{s^{i}_{m, \omega}}^* \bigg)  e^{- i \pqty{m \theta + \omega t}} \bigg] \label{eq_fft_m}
      \end{align}
Again, now that another transform has been completed, the superscripts no longer denote the real component of a complex variable as they are themselves complex. To then understand the relationship between the values in the parentheses, we replace them as $s^r_{m, \omega} = a +i b$ and $s^i_{m, \omega} = c + id $. Doing so gives 
\begin{alignat}{4}
     &s_{m,\omega} &&= s^r_{m, \omega} +i s^i_{m, \omega}  &&= (a-d) + i(b+c) \\
    &s_{m,-\omega} &&=  \pqty{s^{r}_{m, \omega}}^* +i \pqty{s^{i}_{m, \omega}}^*  &&= (a+d) - i (b-c) \\
    &s_{-m,\omega} &&= s^r_{m, \omega} - i s^i_{m, \omega} &&= (a+d) + i (b-c) \\
    &s_{-m,-\omega} &&= \pqty{s^{r}_{m, \omega}}^* - i \pqty{s^{i}_{m, \omega}}^* &&= (a-d) - i (b+c)
\end{alignat}
These relations reveal that 
\begin{align}
    s_{m,\omega}  = s_{-m,-\omega}^* \hspace{1cm} \mathrm{and} \hspace{1cm} s_{m,-\omega}  = s_{-m,\omega}^*.
\end{align}

\bibliographystyle{jfm}
\bibliography{jfmbib}

\end{document}